\documentclass[aps,twocolumn,raggedbottom,prb,showpacs,nobalancelastpage,amssymb,groupedaddress]{revtex4}
\usepackage[english]{babel}
\usepackage{graphicx}
\usepackage{amssymb}
\usepackage{amsmath}
\usepackage{amscd}
\usepackage{eucal}
\usepackage{color}
\usepackage{bm}
\usepackage{upgreek}
\usepackage{subfigure}

\def\be{\begin{equation}}
\def\ee{\end{equation}}
\def\bea{\begin{eqnarray}}
\def\eea{\end{eqnarray}}
\def\bsub{\begin{subequations}}
\def\esub{\end{subequations}}

\def\u{\uparrow}
\def\d{\downarrow}

\newcommand{\proba}[3]{ {\bf P}^{\rm #1}_{#2#3}}

\newcommand{\src}[1]{ {\pmb \mu}^{\rm src}_{#1}}
\newcommand{\currsrc}[1]{ {\bf j}^{\rm src}_{#1}}

\newcommand{\probasc}[4]{P^{\rm #1}_{#2#3,#4}}

\newcommand{\curr}[2]{ {\bf j}_{#1#2}}

\newcommand{\pot}[2]{ {\pmb \mu}_{#1#2}}

\def \Rs{{\cal R}_{\rm sh}}

\def\le{\left}
\def\rg{\right}

\begin{document}

\title{Mesoscopic Current-In-Plane Giant Magneto-Resistance}

\newcommand{\spsms}{SPSMS, UMR-E 9001 CEA / UJF-Grenoble 1, INAC, Grenoble, F-38054, France}
\newcommand{\spintec}{SPINTEC, UMR 8191  CEA / CNRS / UJF-Grenoble 1 / Grenoble-INP, INAC, Grenoble, F-38054, France}
\newcommand{\lps}{Laboratoire de Physique des Solides, CNRS, Universit\'{e} Paris Sud, UMR 8502, B\^{a}timent 510, F-91405 Orsay Cedex, France}

\date{\today}

\author{Cyril Petitjean}
\affiliation{\spsms}
\author{Mairbek Chshiev}
\affiliation{\spintec}
\author{Jacques Miltat}
\affiliation{\lps}
\author{Xavier Waintal}
\affiliation{\spsms}

\date{\today}

\begin{abstract}
We develop a three dimensional semiclassical theory which generalizes  the Valet-Fert model  in order to account for non-collinear systems with magnetic texture, including e.g. domain walls or magnetic vortices. The theory allows for spin  transverse to the magnetization to penetrate inside the ferromagnet over a finite length
and properly accounts for the Sharvin resistances. For ferromagnetic-normal-ferromagnetic multilayers where the current is injected in the plane of the layers (CIP), we predict the existence of a non zero mesoscopic CIP Giant Magneto-Resistance (GMR) at the diffusive level. This mesoscopic CIP-GMR, which adds to the usual ballistic contributions, has a non monotonic spatial variation and is reminiscent of conductance quantization in the layers. Furthermore, we study the spin transfer torque in spin valve nanopillars. We find that when the magnetization direction is non uniform inside the free layer, the spin torque changes very significantly and simple one-dimensional calculations cease to be reliable. 
\end{abstract}


\maketitle

Quantum effects in electronic transport are usually small deviations to the   classical Ohm's law. Famous examples include
weak localization corrections~\cite{Anderson:1979,Gorkov:1979} and universal conductance fluctuations~\cite{Altshuler:1985,Lee:1985} which can be observed in diffusive systems. On the other hand, in small mesoscopic systems quantum mechanics can lead to an entirely different behavior of the conductance. Perhaps the most paradoxical example is the case of a smooth nanoconstriction in a two dimensional electron gas (quantum point contact) where by varying the strength of the confinement the conductance variation has a step-like character with plateaus quantized in unit of  $2 e^2/h$~\cite{Wees:1988, Wharam:1988}. 
This observation was at first a bit puzzling since the quantum point contact does not have any source of scattering, neither elastic nor inelastic. The crucial concept for clarifying the picture was the notion of reservoirs (electrodes) attached to the quantum point contact where the energy relaxation takes place. 

We revisit  bellow this issue in the context of the Giant Magnetoresistance (GMR) effect~\cite{Baibich:1988,Binasch:1989} observed in magnetic multilayers comprising ferromagnetic layers (${\rm F}$) separated by a normal metal (${\rm N}$) structures. We consider ${\rm F|N|F }$ trilayer spin valve structure. There exist two geometries for GMR.  In the CPP geometry one injects the current perpendicular to the plane~\cite{Pratt:1991,Bass:1999,Piraux:1994} of the layers. The CPP-GMR can be understood simply within a two-current model where electrons with up and down spins experience different resistances as they cross the two magnetic layers. As a result, the configuration where the magnetizations of the  FM layers are parallel (P) has a different resistance from the  anti-parallel (AP) one, hence giving rise to the GMR. However, the original experiments~\cite{Baibich:1988,Binasch:1989,Parkin:1990,Parkin:1995} were performed in CIP geometry where one injects the current within the plane (CIP) of the layers. The CIP setup is much simpler experimentally but the two current model (as well as its generalization, the Valet-Fert~\cite{Valet:1993} drift diffusion theory) predicts a vanishing GMR. This usually complicates the interpretation of the experiments as one
needs to introduce (sub mean free path quantum) microscopic approaches and the resulting GMR can be quite sensitive to the details of the model~(~\cite{Camley:1989, Levy:1990,Vedyayev:1993,Vedyayev:1997b, Vedyayev:1998a} and references therein).
Here, we predict the existence of an additional contribution to CIP-GMR which already exists at the drift-diffusion level. This contribution has the same origin as the quantification of conductance and should dominate the usual ballistic contributions for mesoscopic systems.

In this work we address the following problems. First,  a physical explanation for the role of conductance quantization
in CIP-GMR is provided. Next, we develop a theoretical framework allowing for a quantitative prediction of this effect. This framework which we refer to as Continuous Random Matrix Theory in $3$ Dimensions (CRMT3D) extends on a previous one dimensional semi-classical approach CRMT and goes beyond existing models~\cite{Strelkov:2010}.
Finally, we apply CRMT3D to the study of the spin transfer torque effect in a CPP nanopillar.

\section{Sharvin resistance and CIP-GMR: physical picture.  }
A very transparent way to describe transport in a quantum system is the Landauer-B\"{u}Ÿttiker formalism~\cite{Buttiker:1986} where one describes
quantum transport with a scattering matrix that relates the amplitudes of outgoing  to the incoming modes.
The quantum system is treated as a waveguide with $N_{\rm ch}\sim A/\lambda_F^2$ propagating modes. In the metallic systems considered here, their cross section $A$ is much larger than the square of the Fermi wave length $\lambda_F$, so they typically contain thousands of conducting channels. The Landauer formula relates the transmission probability $T_n$
of channel $n$ with the conductance as $ g= {e^2/h} \sum_{n=1}^{N_{\rm ch}} T_n$ so that for a perfectly transparent system ($T_n=1$) the system has a finite Sharvin resistance~\cite{Sharvin:1965} $\Rs = h / (N_{\rm ch} e^2)$ . As mentioned above, the Sharvin resistance  (also known as contact resistance) has been observed repeatedly, in particular in semiconductor based mesoscopic systems where $N_{\rm ch}$ can be tuned with the help of an electric field. 
The Sharvin resistance needs to be accounted for only once: when two systems $A$ and $B$ (of resistances ${\cal R}_{A}$ and ${\cal R}_{B}$) are connected in series, the total resistance is given (in the many channels limit  $N_{\rm ch}\gg 1$ considered here) by\cite{datta,Rychkov:2009} ${\cal R}_{AB} = {\cal R}_{A} + {\cal R}_{B}  -\Rs$. In other words one adds up the intrinsic resistances ${\cal R}_{A}-\Rs$ of the conductors in series (regular Ohm's law) and adds {\it once} a contact $\Rs/2$ resistor for each electrode.
For metallic spin valves under consideration here, the Sharvin resistance is the leading quantum correction to Ohm's law. $\Rs$ typically corresponds to the resistance of an interface between two different metals or, say, 10 nm of bulk material. This resistance would normally be difficult to distinguish from a series resistance coming from the measuring apparatus. However, we shall see that in the case of the CIP geometry, the GMR signal simply vanishes at the purely classical level (more precisely in the  limit of validity of the Valet-Fert model~\cite{Valet:1993} described below) and the presence of the Sharvin (quantum) resistance provides the leading source of GMR of mesoscopic samples.
\begin{figure}[h]
\begin{center}
\rotatebox{0}{\resizebox{8cm}{!}{\includegraphics{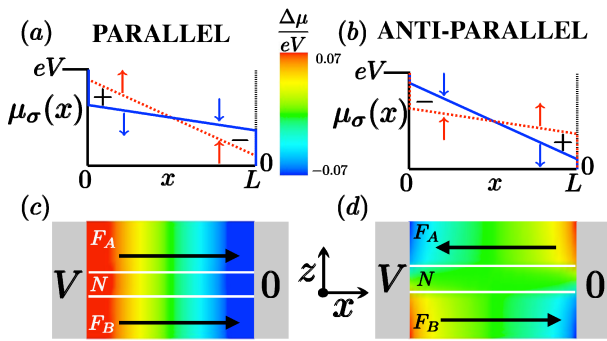}}}
\caption{ Upper panels: Cartoon of the spin resolved chemical potential (up spins: dotted line, down spins: full line) of the top layer of a CIP-GMR setup as a function of the position $x$ in between the electrodes. The chemical potentials drop from $eV$ ($x=0$) to $0$ ($x=L$). The initial and final drop at $x=0/L$ is due to the presence of the Sharvin resistances. Lower panels: numerical simulations of spin accumulation in a ${\rm Co_{3}|Ag_{1}|Co_{3}}$ trilayer with a size $L=20 nm$.  Left (a,c) and Right (b,d) panels correspond respectively to the parallel and anti parallel configurations.   
}
\label{figacc}
\end{center}
\end{figure}

The proper generalization of Ohm's law to a three dimensional magnetic multilayer stack is given by the Valet-Fert 
equations: 
\bea \label{eq:VF3d}
{\bf j}_{\sigma} &=&  - \frac{1}{e \rho_{\sigma}} \pmb{\nabla}  \pmb{ \mu }_ {\sigma} \label{eq:VF3d-ohm}\\
\pmb{\nabla} \cdot{\bf j}_{\sigma} &=& \frac{1}{e \rho_{\sigma}\ell_{\rm sf}^2} \le[\pmb{ \mu }_ {-\sigma}  -\pmb{ \mu }_{\sigma}\rg]\label{eq:VF3d-currcons},
\eea
\noindent where  Eq.(\ref{eq:VF3d-ohm}) is Ohm's law relating the spin  resolved current ${\bf j}_{\sigma}$  to the gradient of the  spin resolved chemical potential $\pmb{ \mu }_{\sigma}$ with  the spin dependent resistivity $\rho_{\sigma}$.  Eq.~(\ref{eq:VF3d-currcons}) expresses the (lack of) conservation of spin current: the divergence of spin 
current is balanced by spin relaxation which is proportional to the spin  accumulation $\pmb{\Delta \mu} = \pmb{ \mu }_ {\uparrow}  -\pmb{ \mu }_{\downarrow}$ and controlled by
the spin diffusive length $\ell_{\rm sf}$.

A cartoon of the system is presented in the lower panel of Fig.~\ref{figacc}: it consists of two magnetic layers 
$F_A$ and $F_B$ separated by a normal spacer $N$ and connected {\it sideways} to the two electrodes to which a voltage $V$ is applied.
To elucidate the role of Sharvin resistances in CIP-GMR, let us first ignore the role of spin-flip processes.
In the absence of Sharvin resistances, one finds that the spin dependence of the resistivity in the magnetic layers is essentially irrelevant: the chemical potential must drop linearly from $eV$ at $x=0$ to $0$ at
$x=L$ irrespectively of the values of the resistivities $\rho_\sigma$. As a result, $\pmb{ \mu }_\sigma(x,z)$
is constant along the growth $z$ direction and there is no current flow along $z$. There is neither spin accumulation in the system nor GMR. The same conclusion can be drawn from a full analysis of the Valet-Fert equations. 
The situation changes drastically when the system is connected  in series with its contact (Sharvin) resistances as schematically sketched in Fig.~\ref{figacc} (a) and (b):
for an electron species (say majority electron) with low resistivity ($\rho_\sigma L\ll \Rs$) the chemical potential drops mostly at the contacts
while for an electron species (say minority electron) with large resistivity ($\rho_\sigma L\gg \Rs$) most of the drop takes place in the bulk.
As a result some spin accumulation builds up in the system. In particular, in the AP configuration, the up spin (for example) chemical potential varies along the $z$ direction leading thereby to some spin current flow along the $z$ axis. The current patterns become different for P and AP configurations and the GMR
is restored. The color code of the lower panels of Fig.~\ref{figacc} represents the spin accumulation profile calculated $ \pmb{\Delta \mu}$  for a ${\rm Co_{3}|Ag_{1}|Co_{3} }$ (thicknesses in $nm$) trilayer using the theoretical approach described in the following section. We observe a clear non  zero spin  accumulation in  Fig.~\ref{figacc}~(c)~and~(d) for  a CIP geometry.  This  effect is  a direct consequence of the presence of  non negligible quantum contact resistances. 

\section{3D Semi-classical theory : CRMT3D.}\label{crmt3d}
%
\begin{figure}[h]
\begin{center}
\rotatebox{0}{\resizebox{8cm}{!}{\includegraphics{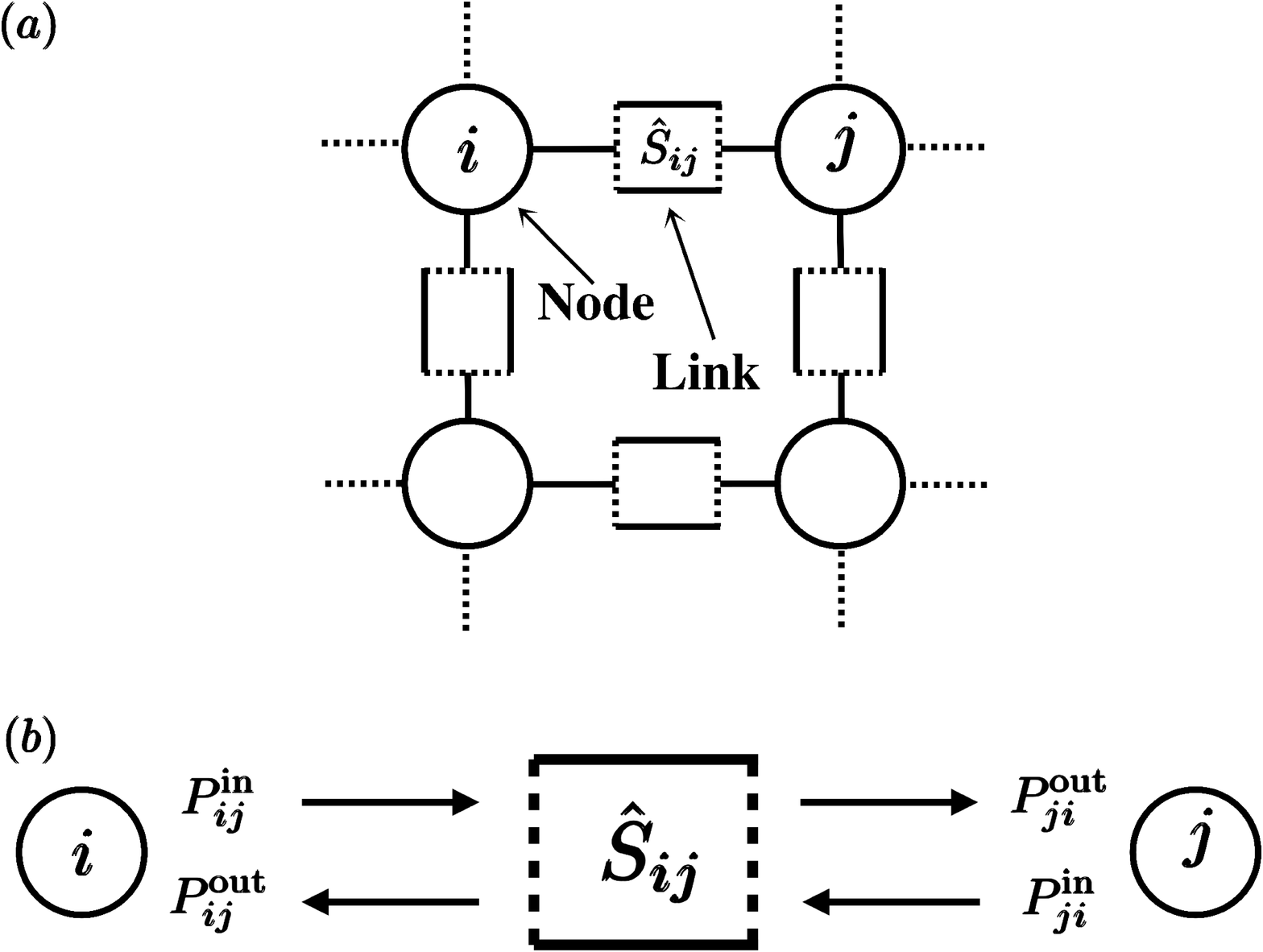}}}
\caption{
Panel (a)  is a cartoon of  discretization of the systems in nodes connected by links.  Each node is labeled by a latin index ($i$, $j$,...) and a link by the two  latin indices of the associated nodes. 
Panel (b)  is a cartoon of  the Scattering matrix $\hat{S}_{ij} $.  We define the region $i$ on the left and $j$ on the right. The  ${\rm in}$ (${\rm out}$) stands for incoming (outgoing) probabilities mode $\proba{in/out}{i}{j}$, where the first index correspond to the side on which the probability current is defined. 
}
\label{figCT}
\end{center}
\end{figure}
In order to provide a quantitative description of aforementioned effects and to capture situations where the magnetization has a non-trivial texture (domain walls, vortices...),  we develop a full 3D semi-classical theory hereafter referred to  as CRMT3D. This approach can be viewed as a generalization of the 3D Valet-Fert theory that properly accounts for Sharvin resistances and non-collinear situations. In addition, CRMT3D can also be considered as a continuous version of the generalized circuit theory~\cite{Bauer:2003b} or equivalently of the random matrix theory developed in~Ref.~[\onlinecite{Waintal:2000}]. CRMT3D is a straightforward generalization of the recently developed CRMT  (Continuous Random Matrix Theory) for unidimensional systems~\cite{Rychkov:2009,Borlenghi:2011}. 
 We refer to Refs.~[\onlinecite{Rychkov:2009, Borlenghi:2011}] for a full derivation of the one dimensional CRMT theory. For completeness,  we recall below the basic objects of the theory before extending it to three dimensions. 
 
A schematic cartoon of the structure of CRMT3D  is presented in Fig.~\ref{figCT}(a) where the system is discretized into many {\it nodes} of small volume. Here we choose a simple Cartesian mesh but this choice is not compulsory. The {\it nodes} are connected to their neighbors by {\it links}. This set of nodes and links forms a circuit theory entirely equivalent
to the so called generalized circuit theory~\cite{Bauer:2003b,Rychkov:2009}. The theory has four basic variables per link $ij$:  $\proba{out}{ i}{ j}$, 
$\proba{in}{ i}{ j}$, $\proba{out}{j}{ i}$ and $\proba{in}{j}{i}$ where the labels in (out) refer to currents going from (to) the nodes while the index order $ij$ ($ji$) indicates that the probability current is defined on node $i$ (node $j$) side as sketched in  Fig.~\ref{figCT}(b). The probability currents 
\be 
\label{eq:P_vector}
\proba{}{i}{j}=
\left(\begin{array}{l}
\probasc{}{i}{ j}{\u} \\
\probasc{}{ i}{ j}{{\rm mx}}\\
\probasc{\, \ast}{i}{ j}{{\rm mx}}\\
\probasc{}{i}{j}{\d}
\end{array}\right)
\ee
\noindent are $4$-vectors that encapsulate the current probabilities for majority ($\probasc{}{i}{ j}{\u} $ ) and minority ($\probasc{}{i}{ j}{\d} $ ) electrons   as
well as spin currents transverse to the magnetization of the layer ($\probasc{}{i}{ j}{{\rm mx}}$).  
The theory is defined  by two fundamental equations relating the outgoing to the incoming currents: one for the links and one for the node. 

The {link equation} is identical to its counterpart in one dimension:
\be\label{eq:hatmat}
 \begin{pmatrix}
\proba{out}{i}{j}  \\ 
\proba{out}{j}{i} 
 \end{pmatrix}= 
 \hat{S}_{ij} 
 \begin{pmatrix}
\proba{in}{i}{j}   \\
\proba{in}{j}{i} 
 \end{pmatrix},
 \ee
where the Scattering matrix $\hat{S}_{ij}$
 \be
\label{eq:hat_s}
\hat{S}_{ij}=
\left(\begin{array}{cc}  \hat r' & \hat t \\ \hat t' & \hat r \end{array}\right),
\ee
is  composed of   $4\times 4$  transmission $\hat t$, $\hat t'$  and reflection $\hat r$, $\hat r'$ material dependent subblocks.
The $\hat{S}_{ij} $ matrix of a thin slice of material of width $b$ is parametrized by two matrices $\Lambda^t$ and $\Lambda^r$:
\be
\label{eq:thin}
\hat{t} = 1 - \Lambda^t \, b, \quad \hat{r} = \Lambda^r \,  b.
\ee
Finally, the matrices $\Lambda^t$ and $\Lambda^r$ of a ferromagnetic metal are parametrized 
by  four independent parameters ($\Gamma_{\u}$, $\Gamma_{\d}$, $\Gamma_{{\rm sf}}$, $\Gamma_{{\rm mx}}$) and read,
\bea
\Lambda^t =&&\!\!\!\!\!\!\!\!
\left(\begin{array}{cccc}  
\Gamma_{\u} +\frac{1}{d}\, \Gamma_{\rm sf} & 0              &  0             & - \frac{1}{d} \,\Gamma_{\rm sf} \\ 
           0                     &\frac{1}{d} \Gamma_{\rm mx}&  0             & 0 \\
           0                     & 0              &\frac{1 }{d} \, \Gamma^{\, \ast}_{\rm mx}& 0 \\
        -\frac{1 }{d} \,\Gamma_{\rm sf}        & 0              & 0              & \Gamma_{\d}+\frac{1}{d} \,\Gamma_{\rm sf} 
\end{array}\right),
\label{eq:lambda_t}\\
\Lambda^r =&&\!\!\!\!\!\!\!\!
\left(\begin{array}{cccc}  
\Gamma_\uparrow -\frac{1}{d}\,\Gamma_{\rm sf} & 0              &   0             & \frac{1}{d}\,\Gamma_{\rm sf} \\ 
           0                                               & 0              &   0             & 0 \\
           0                                               & 0              &   0             & 0 \\
      \frac{1}{d}\,  \Gamma_{\rm sf}                        & 0               &   0             & \Gamma_\downarrow- \frac{1}{d}\, \Gamma_{\rm sf} 
\end{array}\right),
\label{eq:lambda_r}
\eea
 where $d=3$ is the spatial dimension. As  expected, for a  unidimensional system ($d=1$)  Eqs.~(\ref{eq:lambda_t},\ref{eq:lambda_r}) reduce to the  one obtained in the  CRMT approach, Eqs.(36,37) of Ref.~[\onlinecite{Borlenghi:2011}].  The meaning of  the  parameters are also identical and   thus linked to  $5$ physical lengths :  the 
spin resolved mean free path $\ell_{\sigma} = 1/\Gamma_{\sigma}$,  the  spin diffusion length $\ell_{\rm sf} =[4\Gamma_{\rm sf} (\Gamma_\u +\Gamma_\d)]^{-1/2}$,   the transverse spin penetration length $\ell_{\perp}$ and the Larmor precession length $\ell_{\rm L}$. The two latter are  encoded into the complex number $\Gamma_{\rm mx}= 1/\ell_\perp+i/\ell_{\rm L}$.
Note that although the role of $\ell_{\perp}$ and $\ell_{\rm L}$ can lead to interesting physics, in the numerical simulations performed in this paper, we restrict ourself to situations where these lengths are very small (sub nanometer) so that spin torques essentially develop at the normal metal-ferrromagnet interface.
For a normal material,  we substitute in   Eqs.~(\ref{eq:lambda_t}),  $\Gamma_{\sigma} \Rightarrow\Gamma$  and 
  $\frac{1}{d}\Gamma_{\rm mx } \Rightarrow \Gamma +\frac{2}{d}\,\Gamma_{\rm sf}$  so that 
$\Lambda^t$   remains invariant upon arbitrary rotation of the spin quantization axis. $\Lambda^r$  is obtained by following  the same procedure but  with $\Gamma_{\rm sf}$ replaced by $- \Gamma_{\rm sf}$.
The scattering matrices describing the interface between two metals are strictly identical  to those  developed in the one dimensional CRMT case to which
we refer for their expression (See section E of Ref.~\onlinecite{Borlenghi:2011} for details).

Although the natural variables  of the theory are the probabilities $\proba{in/out}{i}{ j}$, they are intrinsically related to the spin  resolved chemical potential
\be \label{eq:mu}
\pot{i}{ j} = \frac{1}{2} \le[ \proba{in}{i}{ j} + \proba{out}{i}{ j} \rg] 
\ee and  the spin  resolved currents
\be\label{eq:j}
\curr{i}{ j}= \frac{1}{e \Rs }  \le[ \proba{in}{ i}{ j} - \proba{out}{ i}{ j}\rg] 
\ee
 flowing from $i$ to $j$. 
The equation at the node is  obtained  by enforcing two conditions. First the chemical potential $\pmb{\mu}_{i}$ depends on $i$ only (and not on the link $ij$):
\be \label{eq:mucst}
\pmb{\mu}_{i}\equiv \pot{i}{ j} , \,  \forall j \in Z_i,
\ee
where  $Z_i$  is the set of neighbors of node $i$. Second, the current is conserved at each node,
\be \label{eq:cons}
\sum_{j \in Z_i}\curr{i}{j} = \currsrc{i},
\ee
where the source term  $\currsrc{i}$ is present only for the nodes connected to electrodes. It is simply given by
\be
\label{lead}
 \currsrc{i} = \frac{1}{e \Rs} \src{i} \quad
\mbox {with,} \quad
\src{i}=
\begin{pmatrix}
eV_a\\
0\\
0\\
eV_a
\end{pmatrix},
\ee
where $V_a$ is the voltage imposed at the electrode $a$.
Rewriting  Eqs.~(\ref{eq:mucst},\ref{eq:cons})  in terms of probabilities and substituting  $\pmb{\mu}_{i}$,  Eq.~(\ref{eq:mu})  yields
the {\it node equation},
\be
\label{eq:prob-ct}
\proba{in }{i}{j} = - \proba{out}{i}{j} +\frac{2}{|Z_i|} \sum_{j \in Z_i} \proba{out}{i}{j} +\frac{2}{|Z_i|} \src{i}  ,
\ee
where $|Z_i|$ is the total coordination number of the node $i$ (counting the connections to other nodes plus the possible presence
of a connected electrode). 
The set of Eq.(\ref{eq:hatmat}), Eq.(\ref{lead}) and Eq.(\ref{eq:prob-ct}) fully defines the theory.
It is equivalent by construction to CRMT for one dimensional case
and one easily verify that
taking the continuous limit of Eqs.~(\ref{eq:mu},\ref{eq:j}) for collinear system, one recovers the VF equations Eqs.~(\ref{eq:VF3d-ohm},\ref{eq:VF3d-currcons}).
CRMT3D can be used in a variety of ways, both analytical and numerical. A very efficient numerical solution (used in the next section for up to a million nodes) 
consists of simply iterating the set of equations  (\ref{eq:hatmat}), (\ref{lead}) and (\ref{eq:prob-ct}) from an arbitrary starting point until until convergence.

\section{Numerical results for CIP-GMR. }

\begin{figure}[h]
\begin{center}
\rotatebox{0}{\resizebox{8cm}{!}{\includegraphics{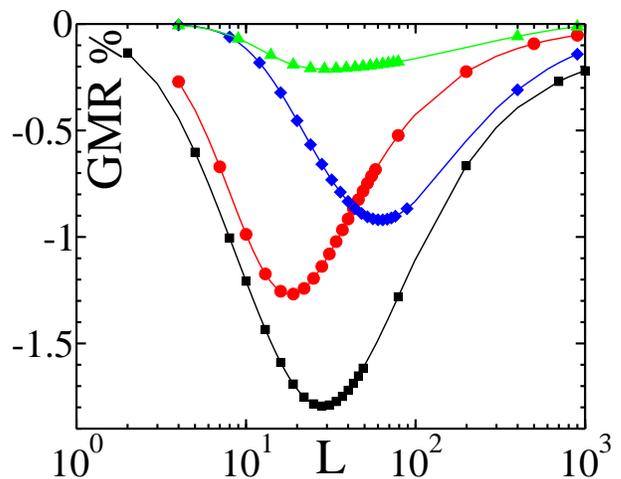}}}
\caption{ Numerical simulations of CIP-GMR (percent) as a function of  the system size $L$ for various ${\rm F|N|F}$ trilayers:  
  ${\rm Co_3|Ag_1|Co_3}$  (Black squares),  ${\rm Py_3|Ag_1|Py_3}$  (red circles),  $ {\rm Py_3 |Cu_1|Py_3}$ (blue diamonds)  and ${\rm Co_3 |Cu_1 |Co_3}$ (green triangles).}
\label{figgmr}
\end{center}
\end{figure}

 We now apply CRMT3D to CIP-GMR. We perform our simulations on various stacks on square samples of size $L\times L$. 
A typical result is presented in Fig.~\ref{figgmr} where the GMR defined as 
GMR$=({\cal R}_{\rm AP} -{\cal R}_{\rm P} )/({\cal R}_{\rm AP} )$ is plotted as a function of $L$ for several  ${\rm F|N|F}$ structures. Here $R_{\rm P}$ and $R_{\rm AP}$ represent the resistance in the $P$ and $AP$ configurations, respectively.
The GMR vanishes in two limiting cases: (i) when $L\ll \min( \ell_{\u},\ell_{\d}  )$, the resistance is entirely dominated by the 
Sharvin resistance which does not depend on spin; (ii) when $L\gg \max( \ell_{\u},\ell_{\d}  )$ the resistance is dominated by the Ohmic
resistance and spin accumulation vanishes as discussed above. Hence, one observes a negative correction for GMR (typically  $-1\%$) for sizes $L\sim  \ell_{\sigma}$ (i.e. when intrinsic and Sharvin resistances have comparable contributions). 
The  actual value of  the GMR  depends on the kind of  material considered (as shown Fig.~\ref{figgmr}) and the various thicknesses of the layers.  For instance, weakly resistive normal metals such as the copper ${\rm Cu}$ (blue squares and green triangles in Fig.~\ref{figgmr})  favors a shunting effect through the spacer which  reduces the mesoscopic GMR signal. 

 In addition, we note the following characteristics: (i) We expect that raising the temperature has opposite effects on the two sides of the negative peak. Indeed, when raising  the temperature the effective mean free paths (which includes both elastic and inelastic scattering) decreases. This makes the Sharvin contribution even less significant for large systems so that the GMR decreases in magnitude. However, for very small systems where the Sharvin contribution dominates, the GMR will start to build up. (ii) The sign of the effect effect is opposite to the CPP case: in the AP configuration, the minority electrons of one layer take advantage of the CIP configuration to propagate more freely in the other layer. (iii) One should keep in mind that this effect occurs  in addition to other microscopic ballistic contributions (typically a few to ten $\%$). Parametrically, this mesoscopic CIP-GMR vanishes algebrically as $1/(1+L/\ell_{\sigma})$ while
ballistic contributions decay exponentially $e^{-l_{\rm N}/\ell_{\sigma}}$ with the width $l_{\rm N}$ of the normal spacer.
It is therefore possible to see the mesoscopic effect only but for most stacks one would observe both effects simultaneously and mesoscopic CIP-GMR  should therefore be observed as a dip in the $L$ dependence of GMR. 

\begin{figure}[h]
\begin{center}
\rotatebox{0}{\resizebox{8.2cm}{!}{\includegraphics{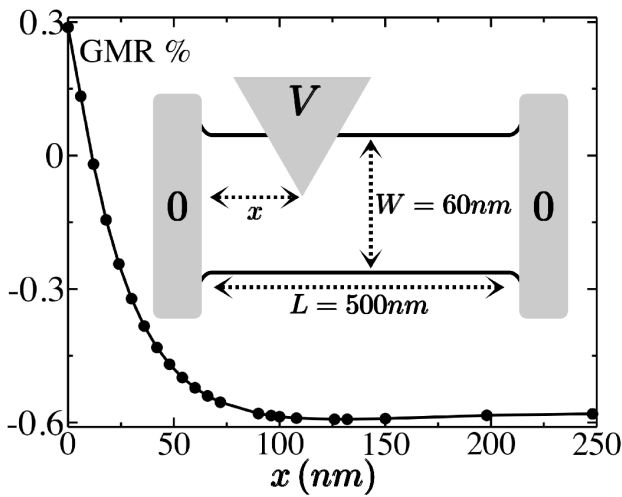}}}
\caption{ mesoscopic GMR in percent as function of  the STM position  $x$ for a  ${\rm F|N|F}$ trilayers:  ${\rm Co_2 |Ag_2| Co_2}$,  
with a  length  $L=500nm$ and a width $W=60nm$. Inset:  Cartoon of  the top view of the setup.}
\label{figgmrcip}
\end{center}
\end{figure}

Measuring the size dependence of CIP-GMR is not an easy experimental task. However, a very similar signature can be obtained by measuring the (two-terminal) GMR using a STM tip as a function of the distance between  the tip and the contact electrode. The setup is presented in the inset of Fig.~\ref{figgmrcip} where the tip is placed on top of the stack at a voltage $V$ while the two other electrodes are grounded. It corresponds to a geometry and sizes very similar to those used in the low temperature STM experiment of Ref.~\cite{LeSueur:2008a}.
As the distance $x$ between the tip and the contact electrode increases, the GMR is anticipated to change from positive (CPP like) to negative when the mesoscopic CIP GMR effects dominates.
%

\section{Conclusion: spin-torque in a realistic CPP spin valve.}
\begin{figure}[h]
\begin{center}
\rotatebox{0}{\resizebox{8cm}{!}{\includegraphics{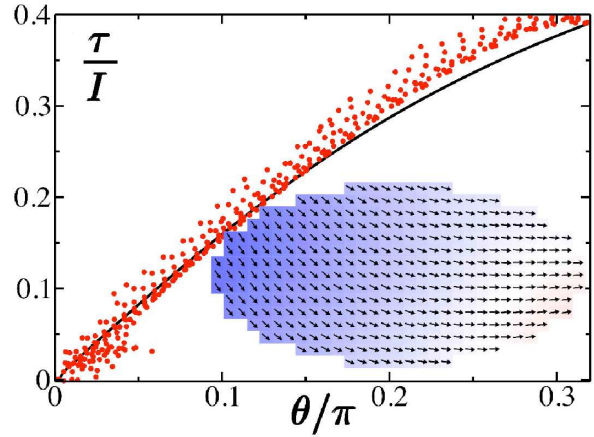}}}
\caption{Torque $\tau$ per total current $I$  as a function of the local angle $\theta$ between the magnetization of the free layer and the reference magnetization of the thick layer for 
a  ${\rm Cu_{48}| IrMn_{8} |Py_{8} |Cu_{4}| Py_{8} |Cu_{20} |Pt_{8} |Cu_{100} }$ pillar for a current density $I=2.5 \,10^7 A.cm^{-2}$.
The full black line corresponds to 1D CRMT calculation  while the  cloud of  red dots 
corresponds to the CRMT3D calculation. Inset: reference magnetic configuration of the free layer, calculated with micro-magnetic simulations~\cite{Miltat:2007}.
}
\label{figtorque}
\end{center}
\end{figure}
To conclude this paper, we take advantage of the capability of CRMT3D to treat systems with magnetic texture and perform a study of spin transfer torque
in a spin valve. We aim at evaluating the role of magnetic texture which is often disregarded in the calculation of spin torque made in micro-magnetic simulations.
Our nanopillar corresponds to the following stack:  ${\rm Cu_{48}| IrMn_{8} |Py_{8} | Cu_{4} |Py_{8} |Cu_{20}| Pt_{8} |Cu_{100} }$. It consists of a polarizing
layer pinned by the ${\rm IrMn}$ antiferromagnetic layer and a free $8\,nm$ permalloy layer. This setup basically corresponds to the experiments reported in Ref.~\cite{Krivorotov:2007} and has been designed in such a way that the current induced magnetization reversal behaves in a coherent way (i.e. as close to a macrospin as possible) so that in this situation the role of magnetic texture is believed to be fairly small. Nevertheless, the Oersted field  which is  present at high current introduces a small "banana shape" magnetic texture, as shown in the inset of Fig.~\ref{figtorque}. Our starting point is the corresponding stationary magnetic configuration of the
free layer obtained from a micromagnetic simulation in presence of the Oesterd field~\cite{Miltat:2007}. In a second step, we perform {\it two} different calculations of
the spin torque: (I) a full CRMT3D calculation the local spin transfer torque $\tau_{I}(x,y)$ in presence of the Banana shape magnetic texture (the polarizing layer which is pinned by the ${\rm IrMn }$ layer is supposed to have no magnetic texture). (II) We take an approach which ignores the role of in plane spin currents: one assumes that current density is homogeneous across the nanopillar and parametrize the local spin transfer  torque $\tau=f(\theta)$ as a function of the angle $\theta$ between the (local) magnetization and the reference fixed polarizing layer. When the system acquires some magnetic texture one uses $\tau_{II}(x,y)\equiv f(\theta(x,y))$. The parametrization $f(\theta)$ is obtained using the one dimensional 
version of CRMT. The effective 1D approach (II) is equivalent to the 3D approach (I) in the absence of magnetic texture and has become quite common in dynamical micromagnetic studies of current induced phenomena~\cite{Slonczewski:2002,Manschot:2004,Xiao:2004,Xiao:2007,Ralph:2008}.
 The results are shown in Fig.~\ref{figtorque} where the effective 1D CRMT approach $\tau=f(\theta)$ (line) is contrasted with the full 3D calculation  where $\tau_{II}(x,y)$ is plotted as a function of $\theta(x,y)$ (red dots). The apparent "noise" of the CRMT3D calculation reflects the fact that the torque {\it is not} a function of $\theta$ only but fully depends on the spatial position $(x,y)$. One can see that even though the general picture is captured by the effective one dimensional approach, a typical error of more than $10\%$ may be observed. We expect that upon performing an integration of the (highly nonlinear) 3D Landau-Lifshitz-Gilbert equation, such a systematic error will result in strong inaccuracy, even in the favorable situation considered here.  We conclude that micromagnetic simulations of real
predictive power, which are highly desirable for spintronic applications,  require to treat magnetic and transport degrees of freedom on equal footing. In particular,
a natural route would be to perform full CRMT3D calculations of the spin transport properties of the device "on the fly" during the micromagnetic simulation.

 \begin{acknowledgments}
We thank T. Valet and P. Brouwer for very useful discussions.
This work was supported by EC Contract No. IST-033749 DynaMax" , CEA NanoSim program,  Nanosciences Foundation (RTRA), CEA Eurotalent and EC  Contract ICT-257159 Macalo.
 \end{acknowledgments}

\end{document}